\documentclass[letterpaper,twocolumn,10pt]{article}
\usepackage{usenix}
\usepackage{epstopdf}
\usepackage{epsfig,endnotes,hyperref}
\usepackage{amsmath}
\begin{document}

\newcommand{\ignore}[1]{}

\ignore{
\addtolength{\parskip}{-5mm}
\addtolength{\dblfloatsep}{-5mm}
\addtolength{\dbltextfloatsep}{-7mm}
\makeatletter
\def\@maketitle{\newpage
 \null
 \setbox\@acmtitlebox\vbox{%
\baselineskip 20pt
\vskip 2em                   
   \begin{center}
    {\ttlfnt \@title\par}       
    \vskip 1.5em                
   \end{center}}
 \dimen0=\ht\@acmtitlebox
 \unvbox\@acmtitlebox
 \ifdim\dimen0<0.0pt\relax\vskip-\dimen0\fi}
\makeatother
}


\title{Mobile Device Identification via \\ Sensor Fingerprinting}


\author{
 {Hristo Bojinov}\\
 Stanford University
 \and
 {Dan Boneh}\\
 Stanford University
 \and
 {Yan Michalevsky}\\
 Stanford University
 \and
 {Gabi Nakibly}\\
 National Research \\ \& Simulation Center, Rafael Ltd.
}
 
\maketitle

\thispagestyle{empty}

\begin{abstract}
We demonstrate how the multitude of sensors on a smartphone can be used to
construct a reliable hardware fingerprint of the phone.  Such a
fingerprint can be used to de-anonymize mobile devices as they connect
to web sites, and as a second factor in identifying legitimate users to
a remote server. We present two implementations: one based on
analyzing the frequency response of the speakerphone-microphone
system, and another based on analyzing device-specific accelerometer
calibration errors. Our accelerometer-based fingerprint is especially
interesting because the accelerometer is accessible via JavaScript
running in a mobile web browser without requesting any permissions or
notifying the user. We present the results of the most extensive sensor
fingerprinting experiment done to date, which measured sensor properties
from over 10,000 mobile devices. We show that the entropy from sensor
fingerprinting is sufficient to uniquely identify a device among
thousands of devices, with low probability of collision.
\end{abstract}

\section{Introduction}

Many Internet services need reliable identifiers to identify repeat
visitors.  The simplest identifier, a Web cookie, works well, but is
unreliable in case users clear cookies, block 3rd party cookies, or
use private browsing mode.  This lead to the development of
stronger identifiers such as supercookies.  The Panopticlick
project~\cite{eckersley2010unique} showed that desktop browsers are
sufficiently different to be identified.  However, the project noted that
mobile browsers, especially on iOS, are too similar for this approach
to work.

The need for robust identifiers is even stronger on mobile devices.
First, as above, Web sites wish to identify repeat visitors.  Second,
cloud-based services who develop mobile applications often need a robust
phone identifier.  Consider the following
scenario: a user installs a cloud-based app and the app installs an
identifier on the device.  Later the user resets the device to its
factory settings thereby deleting the app and its stored identifier.  The
user then re-installs the app and connects to the cloud service.  At
this point the service cannot tell whether it has already seen the
device before. This simple trick may allow a misbehaving user whose account was
blocked to reconnect to the service using a different identity.

More generally, online device
identification is a topic of much interest to advertising
 networks and organizations providing security services. The need to identify
remote peers is always pitted against concerns that identifying information
may be misused. Rapidly evolving mobile technologies pose new challenges to
preserving user privacy and one of our goals in this paper is to explore these
challenges, and inform the design of future mobile device platforms.

To obtain a robust identifier for mobile devices many app developers have
turned to a hardware ID that survives a device reset to factory settings.  A
recent study shows that 8\% of Android apps use the International
Mobile Equipment Identity (IMEI) as a hardware device ID~\cite{dasient}.
This type of practice is frowned upon to the point that Apple disallows
apps who read the iOS Universal Device ID (UDID) 
on their app store~\cite{udid}.

\paragraph{Our contribution.}
We show that the multitude of sensors on a modern
smartphone can be used to build a robust device ID, or
fingerprint, that is independent of the software state and survives a
hard reset.  Our results show a unique device
fingerprint can be computed without accessing traditional hardware identifiers
such as the IMEI or UDID.  Consequently, simply disallowing app access to 
the device UDID is an ineffective privacy policy.

We experiment with fingerprinting using two sensors:
\begin{itemize}
\item {\bf The speakerphone-microphone system:} the fingerprinting system
  uses the speakers to emit a sequence of sounds at different frequencies and records the
  resulting signals using the microphone.  The fingerprint is computed
  by looking at amplitude and frequency distortions in the recorded
  signals.

\item {\bf The accelerometer:} the accelerometer measures forces in each of
  the three dimensions.  Imprecisions in accelerometer calibration
  result in a device-specific scaling and translation of the measured
  values.   By repeatedly querying the accelerometer we estimate
  these calibration errors by solving an optimization problem
  and using the resulting six values (two for each dimension) as a fingerprint.
\end{itemize}
By collecting sensor measurements from over 10,000 mobile devices we show 
that the resulting fingerprints are robust and survive a hard
device reset.  Moreover there is sufficient entropy in the fingerprint
to reliably identify the device among thousands of devices.

Recently, \cite{AccelPrint} proposed a method to fingerprint an accelerometer while it is vibrating (e.g. during an incoming call or message). This method depends on the surface on which the phone lays and the case in which it is enclosed. In contrast, our method is oblivious to these factors. Ref.~\cite{clarkson2012breaking} and \cite{DASAN} proposed methods to fingerprint loudspeakers. Our method differs by allowing to fingerprint the combination of loudspeaker and microphone thus yielding more fingerprint entropy.

\ignore{
Anonymity on the Internet has received much attention in security research. A number of paradigms and platforms have been proposed in attempts to provide users with desirable services without compromising their privacy. For example, the Tor project~\cite{tor} enables anonymous browsing resilient to network attackers, while a recurring theme in web browser, protocol, and policy design has been to allow for complex web application mash-ups while compartmentalizing user data and restricting unauthorized access to it~\cite{framecommunication,donottrack}.

Concurrently with privacy, mobile security has gained prominence, tracking the growth of mobile computing in general. While native mobile APIs have evolved to a point where app design is rife with security caveats~\cite{androidpermissions}, mobile web browsers have started to offer increasing access to the ``bare metal'' in order to enable web applications with increasingly native feel~\cite{getusermedia,browseraccelerometer}.

In this paper we explore the intersection of mobile security and user privacy. Our main thesis is that mobile devices enable unique fingerprinting methods that allow for identification which can be used to breach user privacy and track users' devices. We demonstrate a novel fingerprinting technique based on gathering sensor calibration data; the technique was applied to large groups of identical devices (16 iPhone/iPod Touch devices, 16 Motorola Droid phones) to demonstrate its feasibility and assess the entropy that it provides for identification purposes.
}

\ignore{
The paper is organized as follows.  We begin in Section~\ref{sec:threat} by defining our threat model and then survey existing device identification techniques. Section~\ref{sec:sensors} reviews the sensors available on mobile devices and discusses their tolerances and corresponding calibration parameters. Sections~\ref{sec:audio} and~\ref{sec:accel} present our experiments with the audio and accelerometer modules. Section~\ref{sec:related} covers related prior work and Section~\ref{sec:conclusion} concludes.
}

\section{Threat Model}
\label{sec:threat}

Device identification may be used for both malicious and benevolent purposes. Here we focus on the offensive potential of sensor fingerprinting.

%
%


A {\em malicious website} may wish to track its users without resorting to any browser storage such as cookies, and without triggering any permissions warnings. Each device that comes into contact with the website is fingerprinted and classified as one of the already known devices or as a new one---if its fingerprint differs sufficiently from those in the database. The malicious website is assumed to be accessed by the user for a long enough time interval, during which the user may possibly leave the device unattended or unused. The user does not suspect that fingerprinting is taking place and does nothing aimed specifically at disrupting the process.

\section{Background}
\label{sec:background}

There are a few standard methods for mobile device identification that can be used by a native (but not browser-based) application. These methods use information exposed by the operating system, however all of them are either applicable to only some versions of an OS or the identifying information reported by the OS can be easily changed by the owner of the device thereby evading identification. We now list some of these standard identification methods in the context of the two most popular mobile operating systems: Android and iOS. For each method we detail its most important features and restrictions. The following is primarily based on~\cite{identifying_app_installations}.

\subsection{Android}

{\bf Device ID}: The method getDeviceId of the TelephonyManager class returns the unique ID for a phone, for example, the IMEI for GSM phones and the MEID or ESN for CDMA phones~\cite{TelephonyManager}. However, no such ID exists for mobile device which do not have telephony capabilities.

{\bf MAC address}: One can retrieve the MAC address of one of the device's network interfaces (e.g. WiFi and Bluetooth). However, the owner of a device is able to change the MAC address of the device. For example, see the ``MAC Address Ghost'' application~\cite{MACAddressGhost}.

{\bf Serial number}: The field SERIAL of the Build class contains a hardware serial number, if one is available on the device~\cite{Build}. This field is only available on version 2.3 and later.

{\bf ANDROID\_ID}: The constant ANDROID\_ID of the Settings.Secure class is a 64-bit number that is randomly generated on the device's first boot and remains constant for the lifetime of the device~\cite{Settings.Secure}. However, the value may change if a factory reset is performed.

\subsection{iOS}

{\bf UDID}: The Unique Device Identifier (UDID) is the primary method for the identification of an iOS device. It is retrieved using the uniqueIdentifier property of the UIDevice class. However, since iOS 5 it has been deprecated~\cite{UIDevice}.

{\bf identifierForVendor}: This is a property of the UIDevice which is only available in iOS 6 and later. It is an alphanumeric string that uniquely identifies a device to the applications vendor, i.e., different application vendors will retrieve different identifiers~\cite{UIDevice}. However, the value of this identifier is deleted once the user uninstalls the last application from a particular vendor. If he later reinstalls an application from that vendor, the a new identifier value will be generated.

{\bf advertisingIdentifier}: This is a property of the ASIdentifierManager class which is only available in iOS 6 and later. It is an alphanumeric string unique to each device and is intended to be use by advertisers~\cite{ASIdentifierManager}. However, the value of this identifier is reset once the device is erased by the user.

{\bf MAC address}: As in Android, a MAC address may be forged by the owner of the device.

\section{Use of Sensors for Identification}
\label{sec:sensors}

The key observation behind this work is that different hardware
instances of a particular sensor are quite different, mainly due to
imperfections in the manufacturing and assembly process.  These
variations introduce biases into the sampled data read from the sensor
that are unique to the specific sensor.  Moreover, these variations
are persistent throughout the life of the sensor.  By measuring the
imperfections we can consistently identify devices carrying these
sensors.

We start by defining different types of sensor biases, and then follow
with an overview of commonly encountered sensor types.

\subsection{Common Bias Types}

{\bf Linear bias}: For many sensor types the value of their measurements can be approximated as a linear function of the true value they measure, i.e. $v_m = v_t  S + O$. Here $v_m$ and $v_t$ are the measured and true values, while $S$ and $O$ are the sensitivity and offset of the sensor (in other words $S$ and $O$ are the calibration parameters specific to the sensor). Ideally, the parameter values should be $S=1$ and $O=0$.  Approximate linear bias can be found in accelerometers~\cite{doscher1998accelerometer}, gyroscopes~\cite{baharev2010gyro}, magnetometers~\cite{renaudin2010complete}, and camera pixels~\cite{holst1998ccd}. We note that in most sensors a linear bias is only an approximation of the actual bias. Some sensors also manifest random and quantization noises, while other sensors exhibit cross-dimensional effects, where the measured value in one dimension affects the measurements in the other dimensions. 
Another example of a linear bias is clock drift. The absolute difference between a clock's time and the true time increases linearly as time goes by. Here the clock's sensitivity is also referred to as the clock skew.

{\bf Tolerance}: For some sensors their measurements can not be modeled as a simple linear function of the true value. In such cases the measurements will be within a predetermined range relative to the true value.  For example, the output gain of a microphone for a specific frequency may vary with $\pm 2$db within the actual input power.

{\bf Timing}: In addition to the bias of the actual measurements a sensor may exhibit variance in the time it takes to produce measurements. Sensor data is often gathered when the hardware triggers an interrupt, signaling that there are new readings available. The timing of this interrupt may vary across devices, and then it can be used as part of an identification scheme. Interrupt timing is relatively difficult to access from application code, but still it may be a viable component in a larger fingerprinting scheme.

\subsection{Common Sensor Types}

Our goal is carrying out a comprehensive survey of sensors that are commonly available on mobile devices, designing specific identification techniques wherever possible.

\begin{table}[ht!]
\begin{center}
\begin{tabular}{|c|c|c|}
\hline
Sensor        & Imperfection   & Comment \\
\hline
\hline
Audio         & tolerance gain & Section~\ref{sec:audio} \\
\hline
Accelerometer & linear bias    & Section~\ref{sec:accel}\\
\hline
Gyroscope     & linear bias    & no baseline \\
\hline
Magnetometer  & linear bias    & variability, \\
              &                & hysteresis \\
\hline
Ambient light  & linear bias   & no baseline, \\
              &                & sporadic data \\
\hline
GPS           & clock skew     & not observable \\
\hline
Touch screen  & misalignment   & no baseline\\
\hline
Camera        & pattern noise  & see~\cite{camera_sensor} \\
\hline
\end{tabular}
\end{center}
\caption{Common mobile device sensors along with a description of their imperfection. Non-covered sensors show the reason why use for identification is difficult.}
\label{tab:sensors}
\end{table}

Table~\ref{tab:sensors} lists the different sensors we have looked at. While in theory most sensors have some sort of measurable bias, in practice the defect may not be readily exposed under ``normal'' conditions, consistent with the threat model that we have outlined, or no baseline measurement is available to calculate the bias. 

In this paper we describe two successful attempts for sensor fingerprinting: audio (microphone/speaker) in  Section~\ref{sec:audio} and accelerometer in Section~\ref{sec:accel}. In Appendix~\ref{app:sensors} we sketch some of the difficulties that hindered identification using other sensors listed in Table~\ref{tab:sensors}.

\section{Device Identification via the Microphone}
\label{sec:audio}

\begin{figure}[ht!] 
\begin{center}
\includegraphics[width=80mm]{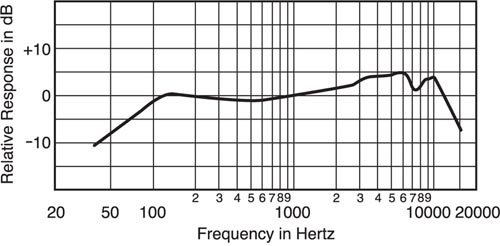}
\caption{A typical frequency response curve for a microphone. Note that for some frequencies the audio is exaggerated (larger than 0 dB) while for others it is attenuated (smaller than 0 dB).}
\label{fig:freq-resp}
\end{center}
\end{figure}

The main specification of a microphone and a loudspeaker is the frequency response graph. A microphone's frequency response is its normalized output gain over a given frequency range. Conversely, a loudspeaker's frequency response is its normalized output audio intensity over a given frequency range. Ideally for both devices, the frequency response should be the same for all frequencies in the range. However, a typical microphone or loudspeaker has a response curve that varies across different frequencies. These variations are dependent on the design of the audio device. Figure~\ref{fig:freq-resp} depicts a typical frequency response curve for a microphone.

Due to manufacturing inconsistencies the frequency responses of each instance of a microphone or a loudspeaker are not identical even if they are of the same model. A device's response for each frequency has a tolerance relative to its response specified by the manufacturer. A typical tolerance for commodity microphone and loudspeakers is $\pm2$db. This frequency response variance gives rise to our first fingerprinting scheme.

\subsection{Fingerprinting Scheme}

\begin{figure}[ht!] 
\begin{center}
\includegraphics[width=40mm]{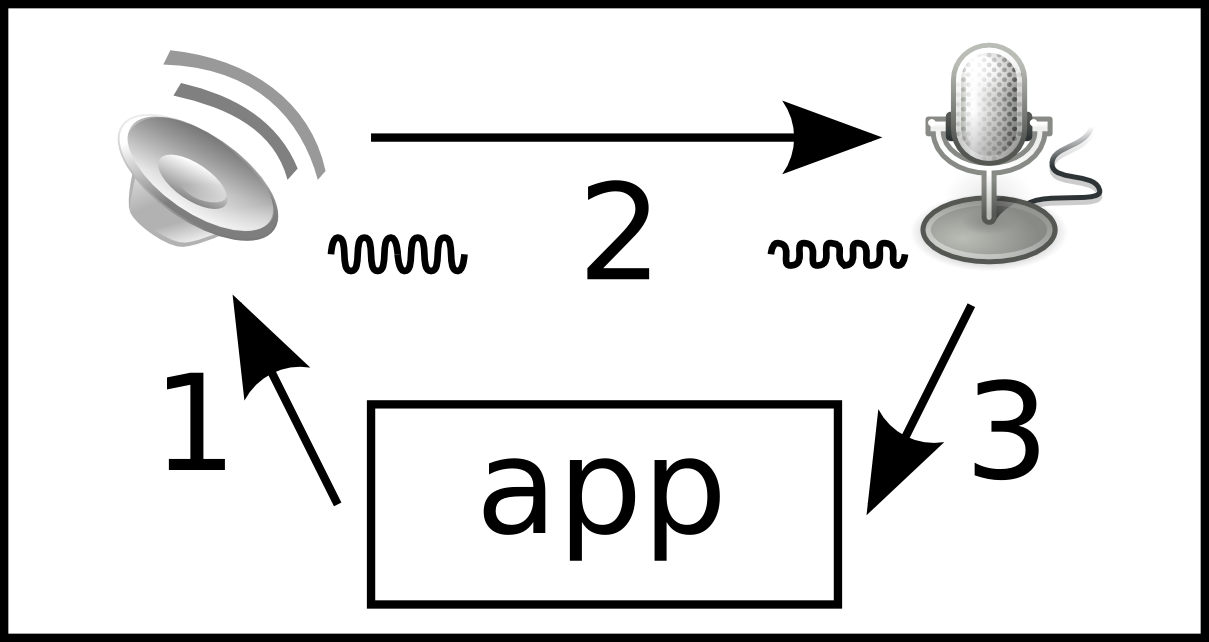}
\caption{Diagram of sound feedback analysis. Within the confines of the device, the application emits sound via the built-in speaker (1), the sound reaches the microphone in a distorted and attenuated form (2), and the application records the microphone signal and analyzes it (3).}
\label{fig:audio_diagram}
\end{center}
\end{figure}

When it comes to fingerprinting, the sound domain is unique because mobile devices have the ability to both transmit (via the speaker) and receive (via the microphone). This in turn allows us to build a completely self-contained fingerprinting scheme, dependent only on the device being in a relatively quiet environment (to minimize signal noise). In our scheme a device's audio fingerprint is the composed frequency response of the device's speaker and microphone. In a nutshell, we play using the speaker an audio signal at a given intensity and we record it using the microphone. We divide the recorded intensity by the original intensity. We refer to this as the \emph{feedback ratio}.  We measure the feedback ratio for several different frequencies. Figure~\ref{fig:audio_diagram} illustrates this process.

The signal recorded by the microphone is processed in the following way. We isolate the main frequency, as well as its harmonics, by computing Fourier coefficients. Specifically, for each played frequency $f_i$, we record 1 second of samples (8000 integers at the typically supported sampling frequency) in the vector $R_i$, and calculate the $j$-th harmonic system response (for $j = 1, 2, \dots$) as follows: 
	\[
	r_{ij} = \sqrt{C(i,j) \cdot R_i + S(i,j) \cdot R_i}
\]
Here $C(i,j)$ and $S(i,j)$ are vectors of 8000 samples of the reference signal as a cosine and sine function at frequency $j f_i$.

\begin{table}[ht!]
\begin{center}
\begin{tabular}{c||c|c|c|c|c|c|c}
\hline
Hz & 220 & 330 & 440 & 550 & 660 & 880 & 1320 \\
\hline
\end{tabular}
\end{center}
\caption{Frequencies at which we measure the feedback ratio for each device. The range and granularity of measurements can be extended, resulting in more data about each device. We have tried to stay below 2000Hz in order to be able to measure at least the second harmonic response at each frequency.}
\label{tab:freqs}
\end{table}

\begin{figure}[ht!] 
\begin{center}
\includegraphics[width=80mm]{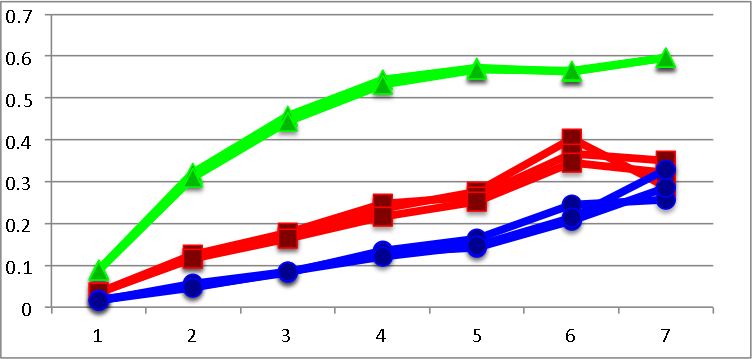}
\caption{Comparison of the first harmonic feedback ratio curves obtained for three devices, each placed in three different locations (the same three locations used by all devices). Each device's curve is labeled with a distinct marker type and color. Feedback ratios (y-axis) are calculated at seven different frequencies (x-axis).}
\label{fig:audio_three_devices}
\end{center}
\end{figure}

We repeat the playback, recording, and analysis at each frequency (Table~\ref{tab:freqs}), obtaining 7 floating-point numbers ($r_{ij}$'s) for each harmonic. Our analysis focused on the first two harmonics ($j=1,2$). Figure~\ref{fig:audio_three_devices} shows a comparison between the first harmonic measurements obtained from three different devices when measured at three different, fixed physical locations each. We note that the feedback ratio for a device is similar but not identical across locations. For example, a device's feedback ratio is dependent on the acoustic properties of surface on which the device lies and the acoustic properties of the device's surroundings. This poses a potential difficulty for this fingerprinting scheme.

\subsection{Experiment: L2-Distance Classification}


\begin{figure}[ht!] 
\begin{center}
\includegraphics[width=80mm]{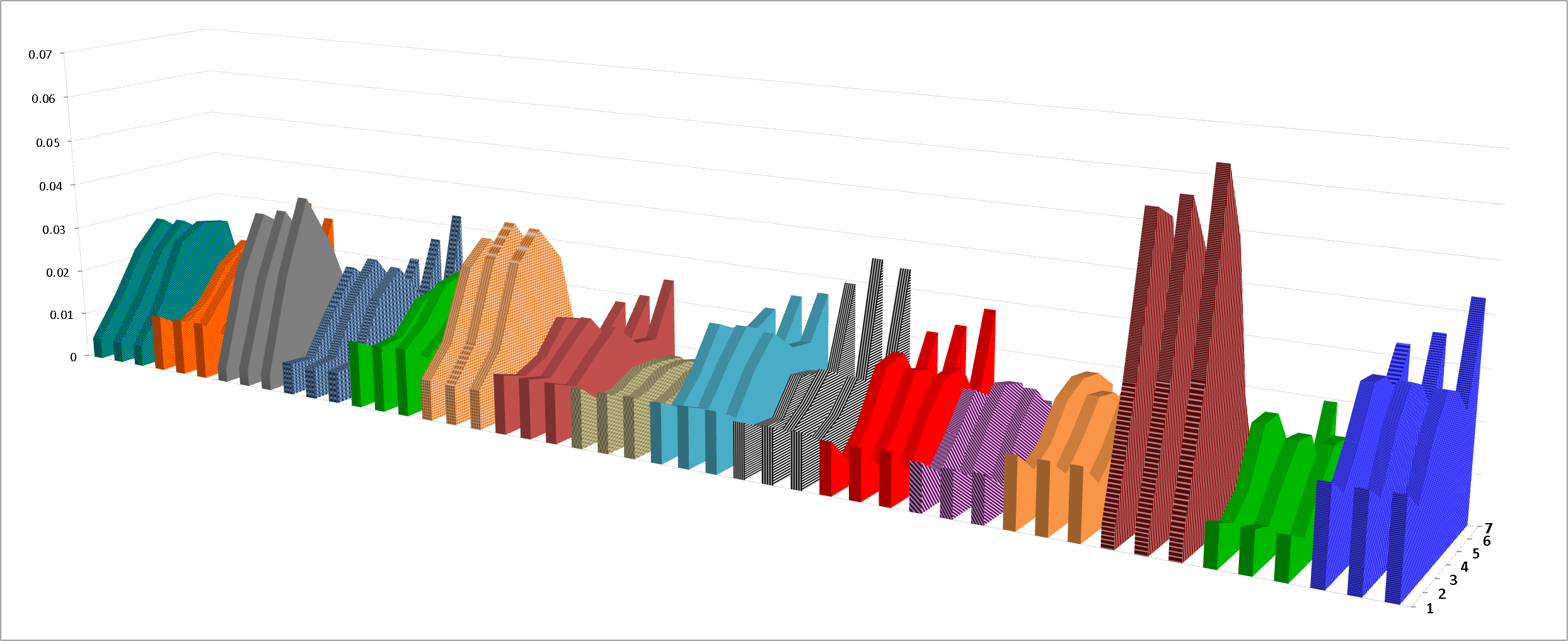}
\caption{Comparison of second-harmonic curves for all devices. Each device is represented by three adjacent curves that have the same color and fill pattern.}
\label{fig:audio_all_devices_h2}
\end{center}
\end{figure}

We used 16 identical Motorola Droid devices to assess the feasibility of a fingerprinting scheme based on sound feedback analysis. We selected three locations (one on a wooden desk, one on a metal filing cabinet, and one on a composite wood windowsill). Each smartphone was placed at each location and measurements were taken. The measurements were performed by an Android application that we side-loaded onto each phone. The application requires only the RECORD\_AUDIO and MODIFY\_AUDIO\_SETTINGS permissions, and once launched it adjusts the sound volume to a fixed, medium level and then plays a three-second sound at each frequency, recording the middle one second back on a separate Java thread. That one-second recording is used as the vector $R$ in subsequent analysis.

Figure~\ref{fig:audio_all_devices_h2} juxtaposes the processed data obtained from all 16 devices. The similarity of the data from the same device across locations is evident, as are the differences between different devices.

\begin{table}[ht!]
\begin{center}
\begin{tabular}{c||c|c|c|c}
\hline
              & \multicolumn{4}{c}{Distance Metric} \\
Test Location\endnote{Location 1 data was used as the training set in this case.} & A       & B     & B'      & B''     \\
\hline
\hline
      2    & 68.8\% & 100\% & 62.5\%  & 56.3\% \\
      3    & 43.8\% & 75\%  & 50\%    & 37.5\%  \\
\hline
\end{tabular}
\end{center}
\caption{Performance of simple L2 distance-based classification of data from locations 2 and 3 (location 1 data were used for the learning step). When first harmonic (Distance A), and second harmonic (Distance B) feedback are compared, second harmonic is almost always more reliable and performs better than first harmonic. Using the second harmonic's first and second derivatives only appears to have a negative overall impact.}
\label{tab:audio_l2}
\end{table}

In order to estimate the amount of information that can be derived from this type of signal metric we designed a simple algorithm which uses one location's measurement from each device to ``learn'' its fingerprint, and then tries to match the data from the remaining two locations to the device for which the learned fingerprint is the closest. If a data set is matched to the device it originated from, we count it as a correct detection (Table~\ref{tab:audio_l2}).

\subsection{Experiment: Maximum-Likelihood Classification}

From the initial set of results it was clear that using the response amplitude at the second harmonic frequency gives the best results, however the classification accuracy clearly left something to be desired. We performed a second experiment with the same batch of Droid devices (excluding one which was faulty\endnote{The device we excluded was not misclassified in the first experiment, so its exclusion did not contribute to the improved results.}).


\begin{figure}[ht!] 
\begin{center}
\includegraphics[width=50mm]{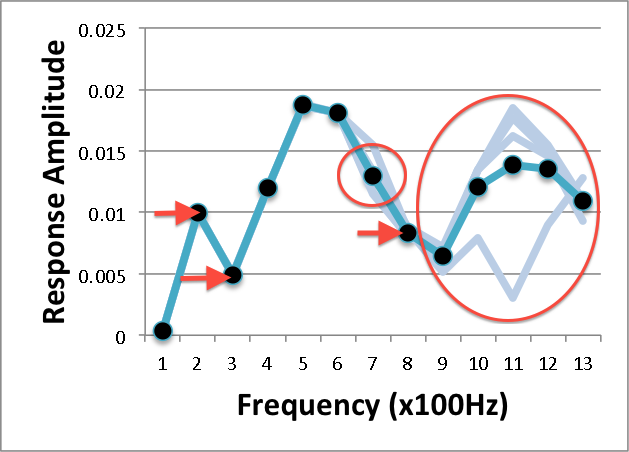}
\caption{Four overlaid frequency response curves at a different location.}
\label{fig:audio_twoloc_2}
\end{center}
\end{figure}

In the second experiment, we measured response amplitudes at 13 different frequencies, from 100Hz to 1300Hz at 100Hz increments. We measured the response of each device at the same three locations we used in the previous experiment, however we ran the measurement four times at each location. Figure~\ref{fig:audio_twoloc_2} illustrates why the simple Euclidean distance-based classification approach was not as precise as we wanted: at some frequencies, on some surfaces, there is simply too much variation---which results in a large penalty when calculating distances. Other frequencies offer much smaller tolerances and can be used to identify the device more accurately.  Fortunately, the {\em maximum-likelihood estimation} method is designed to deal with just this type of situation.

Our new scheme is based on the following simple setup (all frequency response measurements are at the second harmonic frequency---this will be implied for the rest of this section): for each device we assume that its response at a certain frequency is a normally distributed random variable. Given enough samples, we can estimate the mean and variance of the distribution. Later, when we encounter a device, we can calculate the likelihood that it matches any of the known devices in our database---and select the one that maximizes the likelihood. Formally, based on the training data from device $D_k$, at each frequency $f_i$ we estimate the mean of the response $\mu(k,i)$ as well as the variance $\sigma^2(k,i)$. Then we select the device $D_{k_{max}}$ that was most likely to produce this measurement:

\begin{equation}
\label{eq:mle}
k_{max} = \operatorname*{arg\,max}_k \sum_{i=1}^{13} -\frac{(v_i-\mu(k,i))^2}{\sigma^2(k,i)}
\end{equation}

Equation~\ref{eq:mle} follows from the formula for the joint probability of normally distributed, independent variables after taking a logarithm (to turn the product into a sum) and maximizing the resulting sum.

\begin{table}[ht!]
\begin{center}
\begin{tabular}{c||c|c}
\hline
              & \multicolumn{2}{c}{Frequency} \\
Test Location & Main       & $2^{nd}$ Harmonic \\
\hline
\hline
1             & 100\%      & 100\% \\
2             & 100\%      & 100\% \\
3             & 100\%      & 100\% \\
\hline
\hline
1             & 93.3\%     & 100\% \\
2             & 100\%      & 100\% \\
3 (untrained\endnote{The results are similar (symmetric) if another location is omitted from the training data instead.}) & 73.3\%     & 80\% \\
\hline
\end{tabular}
\end{center}
\caption{Performance of MLE-based classification. The first run provides training data from all three locations, then tests at them. The second run omits Location 3 from the training data, yielding worse device recognition rates at the unknown location. The experiment involves 15 devices.}
\label{tab:audio_mle}
\end{table}

The results from using maximum-likelihood estimation are shown in Table~\ref{tab:audio_mle}.

\subsection{Improving Measurement Stealth}
According to the method for microphone and loudspeaker characterization described so far we played each frequency separately. In order to achieve more stealth and shorter sound playback times we examined another method. We choose several frequencies that are not harmonics one of the other, and play them simultaneously. We compute the FFT of the recording and extract the second and third harmonics of the chosen base frequencies. We use the energies corresponding to those frequencies to construct the feature vector corresponding to the device. We obtained feature vectors from 17 devices using this method and tested K-NN classification performance on this data using 10-fold cross validation, achieving correct identification percentage of approximately 95\%.


\section{Device ID Using the Accelerometer}
\label{sec:accel}

An accelerometer measures the acceleration force that is applied to a device along all three physical axes. Recall that the sensor's reading, $v_m$, along the $v$ axis is related to the actual device acceleration, $v_t$, at that axis as follows: $v_m = v_t S_v + O_v$~\cite{doscher1998accelerometer}. Here $S_v$ and $O_v$ are the sensitivity and offset parameters of the accelerometer---note that for a three dimensional sensor there are 6 such parameters. We can use a well known acceleration baseline to measure the accelerometer's offset---Earth's gravity (denoted by $g$)\endnote{Earth's gravity indeed varies a little depending on location, however even these small variations can be predicted.}. At rest the phone experiences an acceleration with a true magnitude of exactly $g$. The orientation of that acceleration depends on the relative orientation of the phone to the Earth's surface.

\subsection{Fingerprinting Scheme}
The accelerometer is convenient to fingerprint for several fundamental reasons: the user often leaves the device still---for instance on a desk, or in a purse; as noted above, when the device is not moving the magnitude of the acceleration vector on the device equals $g$; finally, acceleration can be measured by an Android application that does not require any permissions~\cite{AndroidAccelerometerPermission}, and what is more, iOS as well as Android browsers expose this functionality to websites without notifying the user.

In contrast to audio-based fingerprinting, there is no good way to feed a signal into the accelerometer, namely exert a known acceleration force; instead we take an approach of performing background measurements and waiting until there is enough data to estimate the accelerometer calibration parameters. We perform a measurement every time the phone is at a resting position, or more precisely, the phone is at a constant velocity (no acceleration). Note that in most reasonable cases it is very unlikely that a phone will not be at rest for an extended period of time. Detecting the phone is at rest is relatively straightforward: the measured acceleration vector should be static and its magnitude should roughly\endnote{Note that due to the accelerometer defects we are measuring, it will most likely not be exactly equal to $g$.} be equal to~$g$.  

\subsubsection{Estimating $O_z$ and $S_z$}
Let's assume that throughout our measurements the device will be lying flat and still on a table and thus the $Z$ axis will register practically all the acceleration due to Earth's gravity. If the sensitivity parameter $S_z$ of the sensor is known, then it is easy to estimate the offset from a single measurement $z_m$: $O_z = z_m - g S_z$. Unfortunately, given an arbitrary device $S_z$ is unknown and thus we need two measurements---one with the device facing up ($z_{m^+}$) and one facing down ($z_{m^-}$). Using these two numbers one can calculate the two bias parameters for the $Z$ axis\endnote{For further details see \cite{kionix}.}.
\begin{align*}
	S_z = (z_{m^+} - z_{m^-}) / 2g \\
	O_z = (z_{m^+} + z_{m^-}) / 2
\end{align*}

We will see that this method yields very satisfactory results in a variety of experimental settings even if the surface on which the phone lies is not perfectly level. This method may be exercised without user cooperation since it is quite plausible that the user will sometimes leave their device facing up, and sometimes facing down on a table or a desk. We note that by gathering more data and applying more sophisticated processing, all six accelerometer parameters can be estimated at the same time, with no restrictions on device orientation during measurement (Appendix~\ref{sec:all_six}).

\subsection{Using a Web Page to Profile the Accelerometer}

We started out by building an Android application to track accelerometer readings, filter out any instability, and carry out the simple calculations required to establish $S_z$ and $O_z$ for a given smartphone.

While the application did not require any permissions to access the accelerometer (it is not considered to be a sensitive module by the Android framework), it would need the INTERNET permission to report its findings. We therefore instead built a light-weight JavaScript implementation that runs entirely in the mobile browser.

An implementation contained within a web page has the advantage that the user doesn't need to install anything---assuming the user can be convinced to leave the device facing up and then down for some time, a single visit to a website can result in a fingerprint being calculated.

To collect accelerometer data in JavaScript we implement a function to handle {\tt window.ondevicemotion} events:

{\tt \small
\begin{verbatim}
window.ondevicemotion = function(event) {
  var x = event.accelerationIncludingGravity.x;
  var y = event.accelerationIncludingGravity.y;
  var z = event.accelerationIncludingGravity.z;
  ...
}
\end{verbatim}
}

The first time when our function is called we clear all state and set a text message that instructs the user on how to proceed. Every accelerometer reading results in a new call, where we gather data in batches that are relatively uniform (to avoid noise) and point in a direction that we need (either the positive or negative $Z$ axis, that is, orthogonal to the device screen). When we gather enough positive or negative $Z$ data we instruct the user to flip the device so that we can complete the gathering process.

Finally, when enough data is collected, we estimate $S_z$ and $O_z$ for the device and post the result to our web site for future analysis. We also generate a random number and set a cookie with it in the user's browser. The cookie is also posted to our server to help us correlate different submissions while assessing the error rate for our fingerprinting algorithm.

\subsection{Experiment: Initial Evaluation}

We carried out our first accelerometer profiling experiment on a group of 17 iPhone and iPod Touch devices, obtaining multiple measurements from each.

\begin{figure}[ht!] 
\begin{center}
\includegraphics[width=80mm]{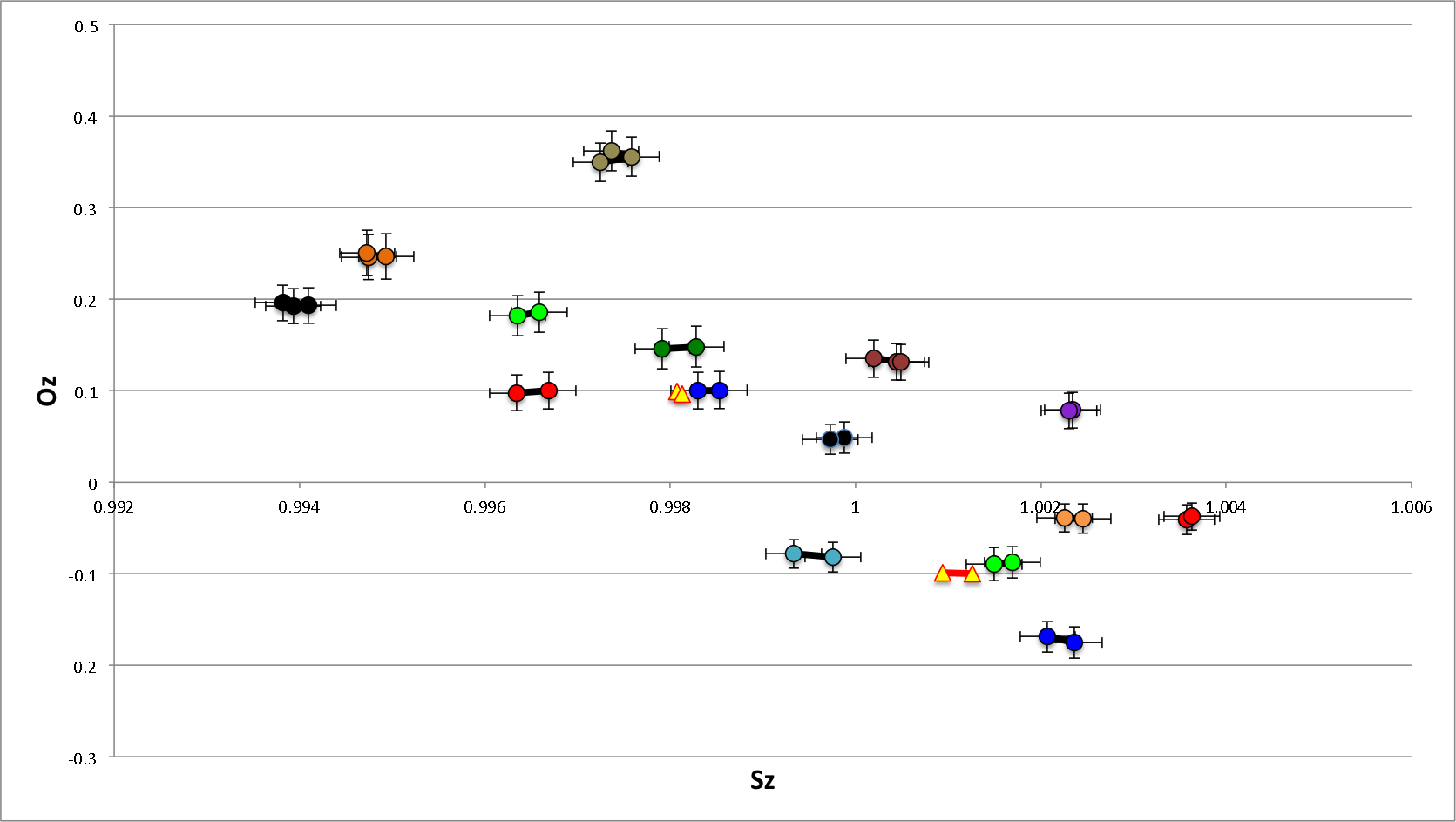}
\caption{Scatter plot of data obtained by accessing the accelerometer profiling web page from 17 iOS devices.}
\label{fig:accel_apple_store}
\end{center}
\end{figure}

Figure~\ref{fig:accel_apple_store} presents a scatter plot of the estimated accelerometer parameters for all devices in the experiment. There were only two pairs of devices whose measurements were ``too close''---one of the devices in each pair uses triangular markers instead of the usual round ones.

In our lab, using the batch of identical Motorola Droid devices from the audio experiments in Section~\ref{sec:audio} we were able to double the amount of data available. Because the Droid devices were too old and did not support JavaScript access to the accelerometer, we implemented an Android application which gathers the necessary data and processes it according to the same algorithm used by the web-based implementation described earlier. The application does not require any special permissions to install, as accelerometer readings are not considered to be significant for preserving user privacy.


The Droid data was also promising: from the 16 devices under study, there was only one near-collision. We combined the data obtained from the two initial accelerometer experiments in order to have the largest possible set of devices and thus assess realistically how feasible it is to identify handsets.

\begin{figure}[ht!] 
\begin{center}
\includegraphics[width=60mm]{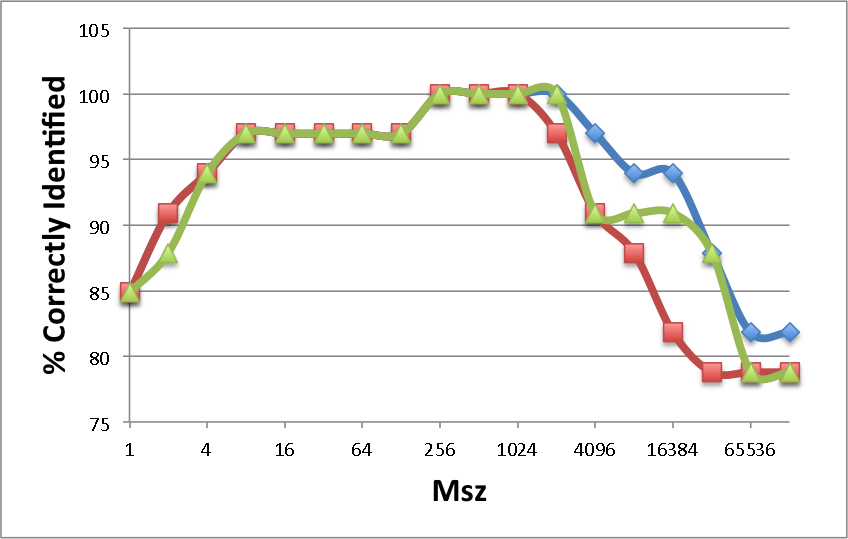}
\caption{Percent correctly identified devices as a function of $M_{S_z}$ in the distance formula. The red, green, and blue lines correspond to different ways of picking the training and test samples for Droid devices---the optimal values for $M_{S_z}$ fall in the same range in all three cases.}
\label{fig:accel_msz}
\end{center}
\end{figure}

Our identification algorithm took two data samples from every device (most iOS devices only had two data points each anyway, due to constraints at the time of gathering; Droids had four samples each and we experimented with picking the first two, last two, and the first and last---Figure~\ref{fig:accel_msz}). We used the first of these samples as the device fingerprint, and the second one as a datapoint to be mapped to the closest of the known fingerprints. We used the square of the Euclidean distance between these points in the plane: $(O_{z_2}-O_{z_1})^2 + M_{S_z} (S_{z_2}-S_{z_1})^2$. Here $M_{S_z}$ is a scaling factor used to reflect the different scale of $O_z$ vs. $S_z$ distances: while $O_z$ typically ranges between $-0.5$ and $0.5$ (Figure~\ref{fig:accel_apple_store}), $S_z$ (being a multiplicative parameter in our model) is much more narrowly spread between $0.99$ and $1.04$. This suggests a scale of about 10 for the values and about 100 for the squares. Indeed, the plot for the rate of correct identification as a function of this scaling factor is shown in Figure~\ref{fig:accel_msz} and confirms this estimate, showing 100\% recognition for $M_{S_z}$ values between about $200$ and $1000$.

\subsection{Experiment: Large-Scale Accelerometer Fingerprinting}
Armed with insight from smaller-scale experiments in accelerometer fingerprinting we set out to expand the scope of the inquiry. Our goal was to gather data from a large number of devices in order to prove that accelerometers can be fingerprinted robustly in the context of a web application

We built an experimental public website~\cite{sensor_id_website} and publicized it in online as well as in printed media. Over the course of two weeks we gathered more than 16,000 submissions.

{\bf Website design.} Our experiment website consists of a data-gathering page, and a dynamically generated chart page that the user can navigate to after submitting his or her device's readings. The chart page displays a scatter plot of all data points that have been recorded so far, and shows the current user's submission in a distinct color (Appendix~\ref{sec:availability}). Effectively, the coordinates of the user's device on the grid comprise the accelerometer's Z-axis fingerprint. 

The user is encouraged to go back to the data-gathering page, repeat the process, and see if the second identifier displayed matches the first one. Our data-gathering web page plants a cookie\endnote{Containing a large random number---a unique ID.} in the user's browser, which makes it possible to correlate data points coming from the same device\endnote{Unless the browser application exited and deleted the cookie; we will ignore this type of scenario here: its presence will strictly degrade our results, so the analysis we report here is conservative.}.

The web site is implemented completely in JavaScript:  no native code or Flash. We found that this method works reliably on a broad range of Android and iOS devices.

\begin{figure}[ht!] 
\begin{center}
\includegraphics[width=60mm]{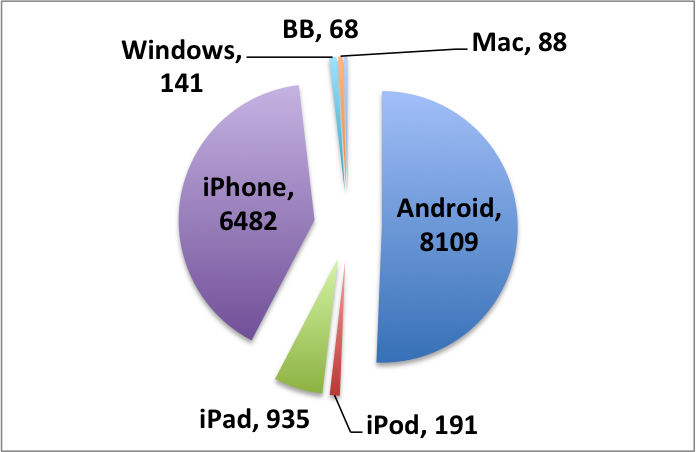}
\caption{Different OS platforms as identified from the User-Agent string in device data submissions. In this chart multiple submissions from the same device are counted as separate entries.}
\label{fig:pie_platforms}
\end{center}
\end{figure}

{\bf Data breakdown.} We looked at several general properties of the data collected in order to ensure it made sense and matches our expectations in terms of quality. In Figure~\ref{fig:pie_platforms}, we show the relative presence of different platforms in our data set. As expected, Android and iOS are the two dominant platforms, with a variety of others present in negligible numbers.

\begin{figure}[ht!] 
\begin{center}
\includegraphics[width=60mm]{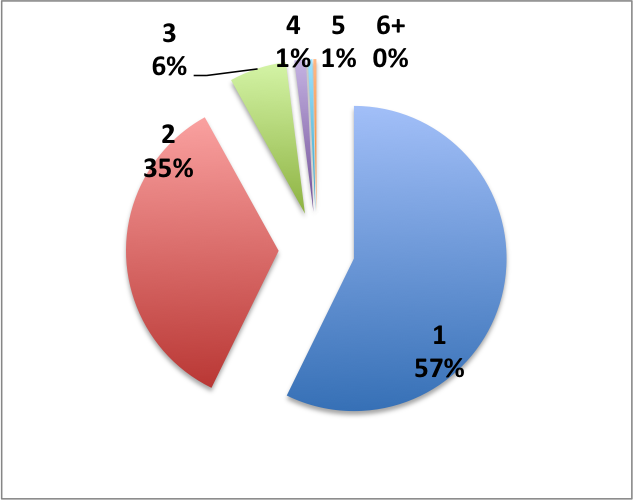}
\caption{Devices grouped by the number of their submissions to our server. This chart reflects actual devices (we have collapsed multiple submissions from the same device into one entry for counting purposes).}
\label{fig:pie_submissions}
\end{center}
\end{figure}

Figure~\ref{fig:pie_submissions} slices the data by the number of submissions we received from a devices. A large number of devices only ran the experiment once, another large group ran it twice (as our web page suggests), with diminishing numbers submitting three, four, and more times.

{\bf Device fingerprinting results.} Our large-scale accelerometer fingerprinting experiment yielded enough data to make a direct calculation feasible. In a pre-processing scan, we create a list of all $O_z$ and $S_z$ distances between the two submissions of two-submission devices. We find that the $95^{th}$ percentile for $O_z$ distance is 0.045 (in other words, 95\% of two-submission devices had a smaller distance than that). For $S_z$ this distance is 0.0037.

When we used these $95^{th}$ percentile distances to divide the $O_z$-$S_z$ scatter plot into blocks of equal size---and counted the data points in each block ($C_{xy}$)---we were able to calculate overall entropy of the distribution as follows:

\begin{equation}
\label{eq:H_direct_Pxy}
P_{xy} = \frac{C_{xy}}{\sum_{x,y} C_{xy}}
\end{equation}

\begin{equation}
\label{eq:H_direct}
H_{direct} = - \sum_{x,y} P_{xy} \log_2 P_{xy} = 7.498
\end{equation}

We verified that small variations of the grid origin had minimal effects on the entropy estimate (specifically, we saw less than $0.01$ bit of entropy difference between the smallest and largest estimated value, $7.493$ vs. $7.502$). We consider this to be a confirmation that the result is robust. 

Note that this entropy measurement is based on the parameters of just a single axis ($Z$-axis). The identifier bits from the accelerometer's $X$ and $Y$ axes are not included in this experiment. If the parameters of these axes can be measured reliably then we would gain additional entropy, allowing the identification of millions of devices.

For identification, we focused on two-submission devices, and asked whether the second data point for such devices is closest to the first data point from the same device, or to some other data point in the set. If the closest data point is from the same device, we count this device as correctly recognized. Note that this is a rather restrictive protocol because we permit all of the data points collected to impact the recognition, including data points from one-submission devices which may or may not have been carefully measured.

In this most restrictive setup, we managed to correctly identify 298 of the 3583 two-submission devices, a success rate of 8.3\%. When we only looked at two-submission devices and further eliminated those whose distance in either $O_z$ or $S_z$ was above the $95^{th}$ percentile, we were able to identify 543 devices, or 15.1\%. While these numbers may look weak, we have to recall that in this identification procedure we are only using the accelerometer, in only one of its dimensions. 

{\bf Improving identification rate via the User-Agent string.}  The
User-Agent string identifies the type of device connecting to a web
site.  The User-Agent data, with no other signals, permitted direct
identification of 544 devices out of the set. The combination of
User-Agent and accelerometer inputs, however, increased the number of correctly
identifiable devices to 1900 devices or 53\%. Removing $95^{th}$
percentile devices further increased the correct recognition rate to~58.7\%.
This shows that the accelerometer fingerprint can be quite effective
at distinguishing devices with identical User-Agent strings.

\subsection{Threat Mitigation}

Device identification via sensor fingerprinting has benign as well as malicious uses. In the context of privacy violation for example, it is worth considering the possible methods for mitigating this threat to mobile users. For any particular sensor, the feasibility of fingerprinting can be practically eliminated by calibrating the sensor at the time of manufacturing. A different, software-only approach can be to add a random value to the sensor output at the OS level. This value can remain constant during continuous use of the device, allowing software such as mobile games to calibrate the sensor if needed. During periods of long inactivity, the random value can change---which would invalidate any device fingerprint that may have been collected already.

We also believe that we have also made a good case for re-evaluating the status of sensor data conferred by browsers and mobile operating systems. Until now, sensor streams such as accelerometer readings have not been considered sensitive information---yet we have demonstrated that they can be used to identify and track devices. As smartphone operating systems and browser technologies mature further, we expect to see more uniform access controls on device sensors.

\section{Related Work}
\label{sec:related}

Sensor fingerprinting has received significant attention in recent years, primarily in the context of de-anonymizing photos by correlating them to images with a known source. In~\cite{camera_sensor}, images taken by different cameras are processed to derive a reference noise pattern that is specific for each sensor. Based on this pattern, additional images are associated with their most likely source. Noise extraction algorithms are a critical part of this approach, and~\cite{camera_enhanced} proposes some further enhancements. The image capture pipeline is investigated in~\cite{camera_stages}, where different stages of the process are revealed to introduce distinct artifacts. These artifacts can be used to design more robust identification algorithms.

Flash (solid state) storage has also been shown to contain unique defects that can be fingerprinted. Both coupling and timing effects are considered by~\cite{flash_fingerprinting}, and shown to yield feasible identification mechanisms. The main difficulty of applying flash fingerprinting to the mobile device domain is the consistent move towards using eMMC-style flash chips~\cite{cooke2007flash}, which hide much of the raw data by building in complex wear-leveling logic.

There are some works that aim to fingerprint a device via the web that go beyond the standard HTTP cookies. Such works are based on software-related features rather than hardware related. Ref.~\cite{eckersley2010unique} showed that parameters of system configuration such as screen resolution, browser plugins and system fonts as well as the contents of HTTP headers -- User-Agent and Accept -- allow to fingerprint a device.  Ref.~\cite{yen2012host} also showed that good device identification can also be achieved using the values of User-Agent, IP address, cookies and login IDs. These values can be achieved using standard logs of web traffic. 

In the past several years it has been shown~\cite{ayenson2011flash} that may web sites identify a web client based on ``super-cookies". These are identifier which are stored on the local host in various persistent ways outside the control of a browser, hence the browser can not impose that standard restriction as of HTTP cookies. 

Some works deal with remote hardware-based fingerprinting. The most well-known example is \cite{kohno2005remote} which showed how to measure a device's clock skew using ICMP and TCP traffic. The clock's skew is shown as a good device identifier. There is also a body of work that propose remote fingerprinting methods based on wireless traffic, for example, radiometric analysis of IEEE 802.11 transmitters~\cite{brik2008wireless}, signal phase identification of bluetooth transmitters~\cite{hall2003detection}, or timing analysis of 802.11 probe request frames~\cite{desmond2008identifying}.

There are a few recent works which independently proposed methods to fingerprint accelerometers and loudspeakers. In~\cite{AccelPrint} is suggested to fingerprint a mobile device using its accelerometer. The proposed fingerprinting method is based on accelerometer output while the phone is vibrating (e.g. during an incoming call or message). Then machine learning algorithms are used to identify a phone based on general features extracted from the accelerometer output, such as mean, std. dev., and skewness. These features are indirectly based on the offset and sensitivity of the accelerometer. This method requires about 30 seconds of accelerometer recording during vibration, which may be hard to obtain if the phone is not set to vibrating mode. Furthermore, the method proposed in~\cite{AccelPrint} is influenced by the surface on which the phone lays and the case in which it is enclosed, while our method is oblivious to these since we fingerprinting the accelerometer while it is at rest.

Ref.~\cite{clarkson2012breaking} and \cite{DASAN} propose to fingerprint loudspeakers. The schemes proposed in these works focus solely on fingerprinting the loudspeakers; in contrast, our method allows to fingerprinting the loudspeaker and microphone combined, thus potentially allowing for more fingerprint entropy. Moreover, our use of the device's microphone removes the need for an external microphone during the fingerprinting process and allows for a more practical attack scenario. Finally, our scheme relies on short synthesized sounds that can be generated at the appropriate timing. Ref.~\cite{DASAN} relies on recording ring-tones and therefore the attacker has to wait for an incoming call (or other event) to trigger the sound.



\section{Conclusions and future work}
\label{sec:conclusion}

We presented a new approach to mobile device identification which
allows for devices to be recognized without relying on soft
identifiers (which may be lost after a device reset). Our
fingerprinting method exploits sensor calibration variations in the
speaker-microphone system and in the accelerometer.
Accelerometer-based identification is particularly noteworthy because
it can be performed by untrusted web code running within a mobile
browser.  We hope that our results illustrate the potential risk of
granting untrusted code access to seemingly benign hardware.

This work raises several interesting open problems.  What other types
of mobile hardware can be leveraged for device fingerprinting?  Can
this be done using imperfections in the baseband processor?  In other
sensors?  How much entropy can be extracted overall and, based on
additional data from a larger set of identical devices, can we obtain a 
high-confidence estimate of the distributions of the measured calibration
parameters?  Is there
sufficient entropy in sensor-based fingerprinting to generate a
hardware-based cryptographic key?  We hope these questions can be
answered by future work.

%

\theendnotes

\bibliographystyle{acm}
\bibliography{sensor_id}

\appendix
\section{Availability}
\label{sec:availability}

\begin{figure}[ht!] 
\begin{center}
\includegraphics[width=40mm]{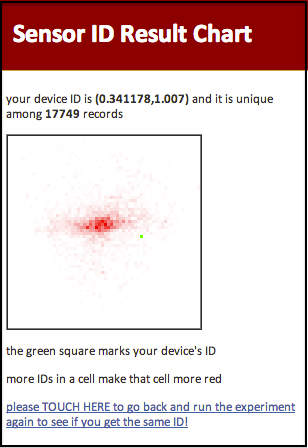}
\caption{Scatter plot of all data gathered, displayed to the user after measurements have been submitted to the server. The green dot depicts the user's device fingerprint. $O_z$ measurements map to the X coordinate, and $S_z$ measurements to the Y coordinate.}
\label{fig:website_chart}
\end{center}
\end{figure}

Mobile devices with compatible web browsers (for example most iOS devices and many Android 4.0+ devices) can be pointed at \url{http://sensor-id.com/} to have their $S_z$ and $O_z$ sensor parameters evaluated. Please follow the instructions and prompts provided. After the website measures your accelerometer parameters, you will be presented with a chart such as the one in Figure~\ref{fig:website_chart}.

The compiled Android application for measuring frequency response ratios can be downloaded from \url{http://sensor-id.com/audio/sensor-id-mic.apk}. In order to run this APK file, the user needs to enable ``non-market applications'' in the Settings application, and then proceed to install either via USB (e.g. via using the {\tt adb install} command) or through another supported mechanism such as a removable SD card. When you run the app, make sure the phone is lying flat on a surface---interference from cables or other objects under the phone is often significant. When fingerprinting is complete, the application appends the frequency responses at the seven frequencies, as well as at their second, third, and fourth harmonics, to the file {\tt /sdcard/mic} and writes out the raw recorded samples to {\tt /sdcard/mic\_dump} (there are 8000 samples at each of the seven frequencies, for a total of 56000 integers, one per line).

\section{Estimating All Six Bias Parameters of the Accelerometer}
\label{sec:all_six}

Using the single-axis ($Z$ only) calibration method one could theoretically estimate the sensor parameters in the $X$ and $Y$ dimensions, however it is very unlikely that a user will position the device on its narrow sides. An alternative approach for estimating the bias parameters for all three dimensions is to gather accelerometer measurements at 6 different and arbitrary resting positions. Note that in this approach although we need to gather measurements for more resting positions we do not assume anything about the device's orientation at those positions. We know that the true acceleration along the $v$ dimension is given by $v_t = \frac{v_m-O_v}{S_v}$. Therefore we have for each measurement $m$ the following equation:
	\[
	\left(\frac{x_m-O_x}{S_x}\right)^2 + \left(\frac{y_m-O_y}{S_y}\right)^2 + \left(\frac{z_m-O_z}{S_z}\right)^2 = g^2
\]

In principle, six such equations (for six different resting positions) would have allowed us to calculate the six unknown bias parameters. However, due to quantization errors and other random noises the above equation should be turned into an inequality of the following form:

\begin{equation*} \label{eq:optimization_constraint}
\left(\frac{x_m-O_x}{S_x}\right)^2 + \left(\frac{y_m-O_y}{S_y}\right)^2 + \left(\frac{z_m-O_z}{S_z}\right)^2 - g^2 \leq \epsilon	
\end{equation*}

where $\epsilon$ is some unknown measurement error. To find the accelerometer deviation parameters in presence of noise we collect multiple measurements $x_m^{\left(i\right)}, y_m^{\left(i\right)}, z_m^{\left(i\right)}$ (as many as it is practical to have) and solve the following optimization problem
\newline
\\
Minimize $\sum_i\epsilon_i^2$ \\
subject to
\begin{equation*}
\tiny{\left(\frac{x_m^{\left(i\right)}-O_x}{S_x}\right)^2+ \left(\frac{y_m^{\left(i\right)}-O_y}{S_y}\right)^2+ \left(\frac{z_m^{\left(i\right)}-O_z}{S_z}\right)^2-g^2=\epsilon_i}
\end{equation*}

where $i$ is the measurement index.



This is not a convex problem in general (depending on the measurements) and therefore we chose to use a numerical gradient descent method to find the parameters $O_x$, $O_y$, $O_z$, $S_x$, $S_y$, $S_z$ that minimize the error. Assuming we have a smart guess for the initial point and given the constraints on the reasonable parameter values we expect that solutions for different sample sets will all converge to the same local minima. We take $O_{3\times1}=0$ and $S_{3\times1}=1$ as the initial point for the algorithm since these are the ideal values from which the device can deviate by a small fraction.

We apply this algorithm to multiple sets of measurements for every device, and obtain labeled samples in a 6-dimensional space. To identify a device given a new set of measurements we repeat the algorithm and obtain an unlabeled sample. We then use nearest neighbor matching (KNN) to associate the sample with a labeled cluster.
Cross-validation of KNN classification over this data yielded a correct classification percentage of 81.3\%.

\subsection{Experiment: Lab Droids in 3D}

We evaluated the algorithm first for 5 devices and then for 16, performing both unsupervised clustering and supervised classification. K-means clustering for the setup of five devices resulted in a perfect identification of samples obtained from the same device. With 16 devices we obtained good clustering, however we did observe some errors. 
Supervised classification with 16 devices yields a correct classification percentage in 81.25\% of the cases, indicating that this method could be a significant contribution to the overall identification process in combination with the other methods.

\section{Difficulties in Identification Using Some Sensors}
\label{app:sensors}

In this section we briefly discuss the difficulties we had in
using certain sensors listed in Table~\ref{tab:sensors} for device fingerprinting.

{\bf Gyroscope}: Measuring the offset and sensitivity of the gyroscope would require subjecting the device to constant angular velocity rotation at different speeds---an experiment that is difficult to carry out even in a lab.
 
\begin{figure}[ht!] 
\begin{center}
\includegraphics[width=60mm]{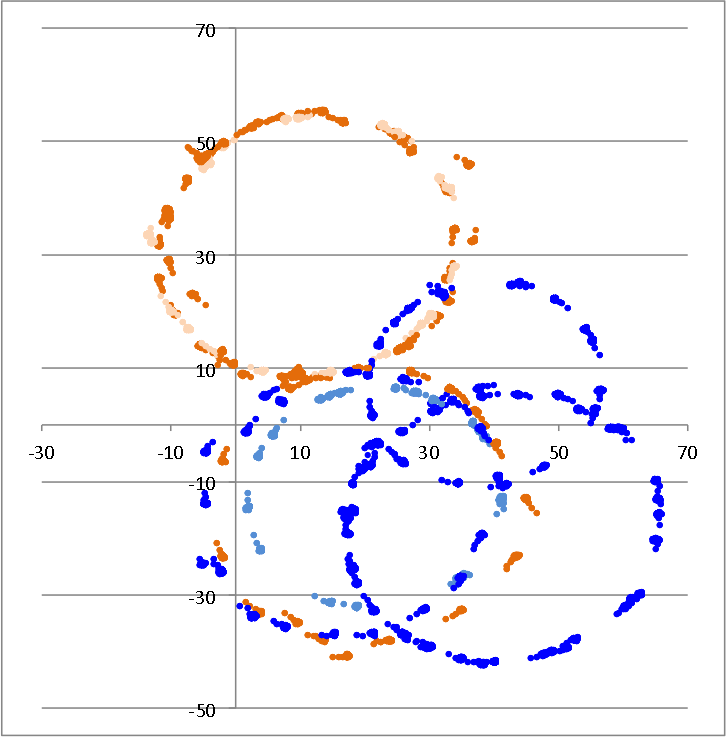}
\caption{Compass (magnetometer) readings taken by placing two different devices (in blue and orange) at the same location on a flat surface, in a variety of orientations. The oval shapes are created due to the varying orientations of the device (with a stable $Z$ component---note that the same device sometimes produces a smaller oval, consistent with a varying $Z$ value across runs, even at the same location). Offset is evident from the offset centers of the ovals, and sensitivity is reflected in the ovals' shapes. Different runs even at the same physical location may produce different results, which demonstrates memory effects that are difficult to correct for.}
\label{fig:compass_issues}
\end{center}
\end{figure}

{\bf Magnetometer}: We carried out some magnetometer experiments which convinced us that although compass readings are a possible source of identification data, the peculiarities of the sensor make practical use next to impossible. Consider for example Figure~\ref{fig:compass_issues}: while sensitivity and offset are evident from the geometry of magnetic field measurements in multiple directions, there are also clear memory effects which can disrupt the estimates. In addition, the variability of the magnetic field can be sometimes significant (e.g. near metallic objects), and sometimes subtle, making corrections difficult and error-prone.

{\bf Light}: Light sensors only provide erratic measurements which are subject to noise due to partial obstruction of the sensor; thus, light measurements are difficult to put in a context that allows for estimation of the imperfections.

{\bf GPS}: A GPS receiver triangulates the location of a phone by calculating its distance to at least 3 GPS satellites. The distances are calculated by measuring the time a signal travels from a satellite to the GPS receiver. The travel time is measured using an inaccurate clock built into the GPS receiver. Previous work~\cite{kohno2005remote} has shown that a clock's skew can identify the clock. However, modern GPS receivers utilize a 4th satellite measurement which allows to take this bias into account. Therefore, the clock's bias does not affect the calculated location.   

{\bf Touch screen}: The touch screen sensor is mounted over the phone's display. Inaccurate assembly process may cause the touch screen to be misaligned with the display. This may cause the user to erroneously tap on locations adjacent to the intended target. This misalignment may serve for identification. However, since it is usually very small and mostly goes unnoticed by the user, it is difficult to measure it. One possible direction to measure it would be, for example, to record the exact locations of the user's taps on the keyboard display. Averaging the tap locations for each key and comparing it against the actual key's location at the keyboard display may allow one to calculate the misalignment. Nonetheless, we expect this method to be highly dependent on the user as much as the touch screen misalignment.       


{\bf Camera}: There are a few works that deal with camera identification using the pixels' bias. The output gain of each pixel is a linear function of the actual intensity of light hitting that pixel. This linear bias is commonly called pattern noise; \cite{camera_sensor} proposes a method to determine a camera's reference pattern noise using a 300 pictures taken by that camera. This serves as a unique fingerprint for the camera. It is shown that this enables to associate with good probability a new picture with the camera that took it. The study was done using 9 cameras. However, no effort was done to assess the expected number of cameras that can be distinguished using this method, and what is more, most of the cameras differ in either model or manufacturer which tends to make identification easier (some evidence is presented however that the results can carry over to identical cameras as well).

\end{document}